\documentclass[twocolumn,trackchanges]{aastex701}
\usepackage{graphicx,url,amssymb,amsmath,color,units,wasysym,epsfig,epstopdf,enumerate, bm, subfigure}

\usepackage[hang,flushmargin]{footmisc}
\usepackage{newtxtext,newtxmath}
\usepackage[T1]{fontenc}
\usepackage{ae,aecompl}
\usepackage[normalem]{ulem}

\newcommand{\inv}{{\text{in}}}
\newcommand{\outv}{{\text{out}}}
\newcommand{\binv}{{\text{bin}}}
\newcommand\sbullet[1][.5]{\mathbin{\vcenter{\hbox{\scalebox{#1}{$\bullet$}}}}}
\newcommand{\BHM}{M_{\sbullet[0.7]}}
\newcommand{\MSUN}{M_{\odot}}
\newcommand{\LSUN}{L_{\odot}}
\newcommand{\RSUN}{R_{\odot}}

\newcommand{\deriv}[2]{\frac{\mathrm{d} #1}{\mathrm{d} #2}}

\newcommand{\bMax}{\text{$\beta_{\mathrm{max}}$}}
\newcommand{\epsGR}{\text{$\varepsilon_{\rm GR}$} }
\newcommand{\epsSA}{\text{$\varepsilon_{\rm SA}$}}

\newcommand{\SPA}{School of Physics and Astronomy, Monash University, Clayton VIC 3800, Australia}
\newcommand{\OzGravMonash}{ARC centre of Excellence for Gravitational Wave Discovery (OzGraV), Clayton VIC 3800, Australia}
\newcommand{\Columbia}{Physics Department, Columbia University, New York, NY 10027, USA}

\begin{document}
\title{Semi-Analytical Model for the Evolution of Stellar Binaries in the Empty Loss Cone of Massive Black Holes}

\author{Samuel McGuire}
    \email{Corresponding author's email address: smcg0013@student.monash.edu}
\affiliation{\SPA}
\affiliation{\OzGravMonash}

\author{Evgeni Grishin}
\email{test}
\affiliation{\SPA}
\affiliation{\OzGravMonash}

\author{Ilya Mandel}
\email{test}
\affiliation{\SPA}
\affiliation{\OzGravMonash}

\author{Yuri Levin}
\email{test}
\affiliation{\Columbia}

\begin{abstract}
Binary star systems orbiting close to a supermassive black hole (SMBH) evolve through encounters with other stars, the SMBH's tidal forces, and the binary's internal dynamics, including general relativistic precession and tides. Many are driven onto highly eccentric inner binary orbits, potentially leading to stellar mergers; other possible outcomes include hypervelocity star ejections or tidal disruption events.  We study the evolution of binaries in the empty loss cone regime, where the outer orbit's angular momentum change per orbit due to scattering off other stars is smaller than the outer angular momentum at the tidal separation radius. We build on the work of Hamers \& Samsing to develop a computationally efficient semi-analytical model that captures the long term evolution of binaries in perturbative regimes where the ratio of the binary tidal separation radius to the pericenter around the SMBH is smaller than 0.15. Crucially, we apply corrections to preserve the orthogonality between the binary's eccentricity and angular momentum vectors, which prevents unphysical eccentricity growth. From these simulations, we find analytical fits for the probability distributions of the final orbital parameters of binaries approaching the SMBH. We find that general relativistic precession efficiently suppresses von-Zeipel-Lidov-Kozai-like eccentricity oscillations and reduces the fraction of merging binaries from $84\%$ with Newtonian physics only, to $3\%$ with precession included. Stellar tides further reduce the merger fraction to $0.4\%$. 
    
\end{abstract}

\section{Introduction}
\label{sec:intro}

The centres of galaxies typically host a Supermassive Black Hole (SMBH) and large stellar clusters around it, leading to unusual stellar evolution with observable signatures \citep{merritt2013}.  
The orbits of stars are scattered due to gravitational interactions within the sphere of influence of the nuclear star cluster \citep{Spitzer1971-bu, Antonini2010-qy}. In particular, the orbital energy and angular momentum around the SMBH can be altered by two-body and resonant relaxation (RR) \citep{RT96, bt_book, k15, baror18, fouvry19, SariFragione2019}. Scattering can lead to extremely eccentric orbits, passing close enough to the SMBH that its tidal forces significantly impact the evolution of the binary orbit.

 A significant fraction of these stars are believed to be in binary systems \citep{Chen2023-yf}, thus it is important to understand the evolution of such systems. For a given inner binary with separation $a_{\rm in}$ and total mass $m_{\rm in}$, it will be tidally separated by the SMBH of mass $M_\bullet$ at a tidal separation distance $R_{\rm TS} = a_{\rm in}(M_\bullet/m_{\rm in})^{1/3}$ from the SMBH. The depth of an inner encounter is often characterised by the parameter $\beta \equiv R_{\rm TS}/R_p$, where $R_p$ is the outer binary’s closest approach to the SMBH. For a deep encounter, $\beta>1$, the tidal forces of the SMBH dominate over the binary's self gravity and disrupt it, leading to many possible outcomes: stellar mergers, hypervelocity stars (HVS) \citep{Hills1988-fp, Huang2021-yg} and tidal disruption events (TDEs) of one \citep{Hills1975-sk, Rossi2021-ax, Gezari2021-ql} or both of the stars \citep{Mandel2015-oe, Yu2024-du}.


In contrast, when $\beta\ll1$, the tidal forces are weaker, and the SMBH alters the inner binary orbit over many outer orbits. This regime can be treated perturbatively, where the combined effects of the von-Zeipel-Lidov-Kozai (ZLK) mechanism \citep{vz191-, lidov62, kozai62, Naoz2016-ku} and successive weak scatterings affect the binary evolution over secular times \citep[see e.g.][and references therein]{huangLu2025}.

The dynamics of binary stars and compact objects around an SMBH has received significant attention \citep[e.g.][]{hopman09, ap12, hoang18, gpf18, fg19, gri25}. The coupled secular ZLK evolution and stochastic perturbations due to other stars could lead to an enhanced gravitational-wave merger rate \citep{hamers_baror18, winter-granic2024}. Rotating or triaxial nuclear clusters can also affect the inner orbit and outer orbit \citep{bub20}. 

Recent works \citep{dodici25, marklund2025, huangLu2025} have explored the evolution of binaries in the inner parsec of a galaxy including the effects from field stars. During high-eccentricity phases of the inner orbit, close passages between the stars can lead to mergers or efficient tidal circularisation, depending on the tidal model. \cite{dodici25} focused on wide binaries on circular outer orbits within the refilling regime, where the ZLK timescale is shorter than relaxation and evaporation timescales, finding that $20-60\%$ of binaries can contract due to efficient chaotic tides. Meanwhile, \cite{marklund2025} simulated a population of binaries near the soft-hard limit and found that about half  ($46\%$) are merging due to ZLK and scattering from field stars.


In this work, we expand upon \cite{Bradnick2017-ed}, who evolved a small sample of $1000$ binaries, undergoing two-body relaxation and tidal interactions from an initial $\beta=0.2$. Because \cite{Bradnick2017-ed} focused on binaries that could lead to TDEs within 1000 outer orbits, their exploration focused on a very specific region of the parameter space. We extend this work to a realistic distribution of initial conditions in a more perturbative regime ($\beta\lesssim0.15$), using a semi-analytical method. Similarly to \cite{huangLu2025}, our method combines analytic prescriptions for the orbital element evolution during the binary's flybys of the SMBH \citep{Hamers2019SecondOrder} with General Relativistic (GR) precession, stellar tides and stochastic two-body relaxation. This semi-analytic approach allows us to simulate populations of $10^5$ binaries for up to $2\times 10^5$ outer orbits, allowing us to statistically characterise the population-level outcomes of binaries in galactic nuclei.

Importantly, our method incorporates corrections to the equations presented in \cite{Hamers2019SecondOrder}, that to our knowledge are not present in \cite{huangLu2025}, in order to preserve the magnitude and orthogonality relations between the eccentricity and angular momentum vectors. These corrections prevent the unphysical eccentricity growth that occurs otherwise, which we find can artificially decrease the binary pericenter distances by orders of magnitude. We further build upon the work of \cite{huangLu2025} by performing a Monte-Carlo simulation of a population of a wide variety of binaries to derive and fit probability distributions for the final eccentricities, separations, and orientations of binaries that approach the SMBH.

Our method does, however, rely on several simplifying assumptions. Notably, we largely neglect the direct impact of individual stellar flybys on the inner binary orbit and employ a simplified tidal model. These approximations introduce systematic uncertainties that must be considered when interpreting our final parameter fits.

Section~\ref{sec:Methods} describes our semi-analytical model including corrected secular evolution, tidal and relativistic effects and two-body relaxation. Section~\ref{sec:singleSysResults} presents the results for a single system. Section~\ref{sec:MonteCarlo} provides the initial conditions and main results of the population study of $3\times 10^5$ binaries. Section~\ref{sec:Discussion} discusses our work in the broader context of recent results, our limitations and future prospects. We conclude in Section \ref{sec:Conclusion}.

\section{Semi-Analytical Model for Simulating Binaries in the Galactic Centre}
\label{sec:Methods}

The evolution of the inner stellar binary orbits on highly eccentric outer orbits is influenced by several factors, as described below. Firstly, as the binary passes near the SMBH, the latter's tidal force changes the eccentricity and the inclination of the inner orbit (while leaving its semimajor axis largely unchanged). Secondly, the outer orbit is perturbed due to stochastic interactions with other stars in the galactic centre. We model the former using the analytical calculation of \cite{Hamers2019SecondOrder} and explore its regime of validity. We also include the additional precession of the argument if pericentre of the inner orbit due to tidal deformation and general-relativistic effects. The latter stochastic interactions are directly modelled as a random walk. 


\begin{figure}
    \centering
    \includegraphics[width=0.95\linewidth]{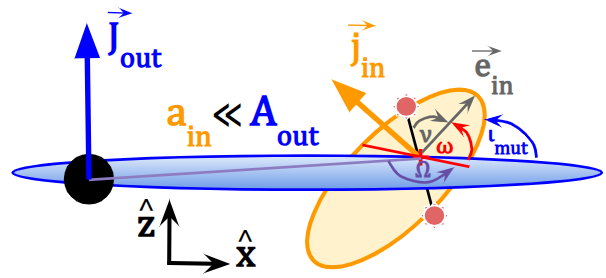}
    \caption{A sketch illustrating the relevant angles and vector quantities; see Section \ref{sec:Methods}.}
    \label{fig:sketch}
\end{figure}

\begin{figure*}[t!]
    \centering
    \includegraphics[width=0.8\textwidth]{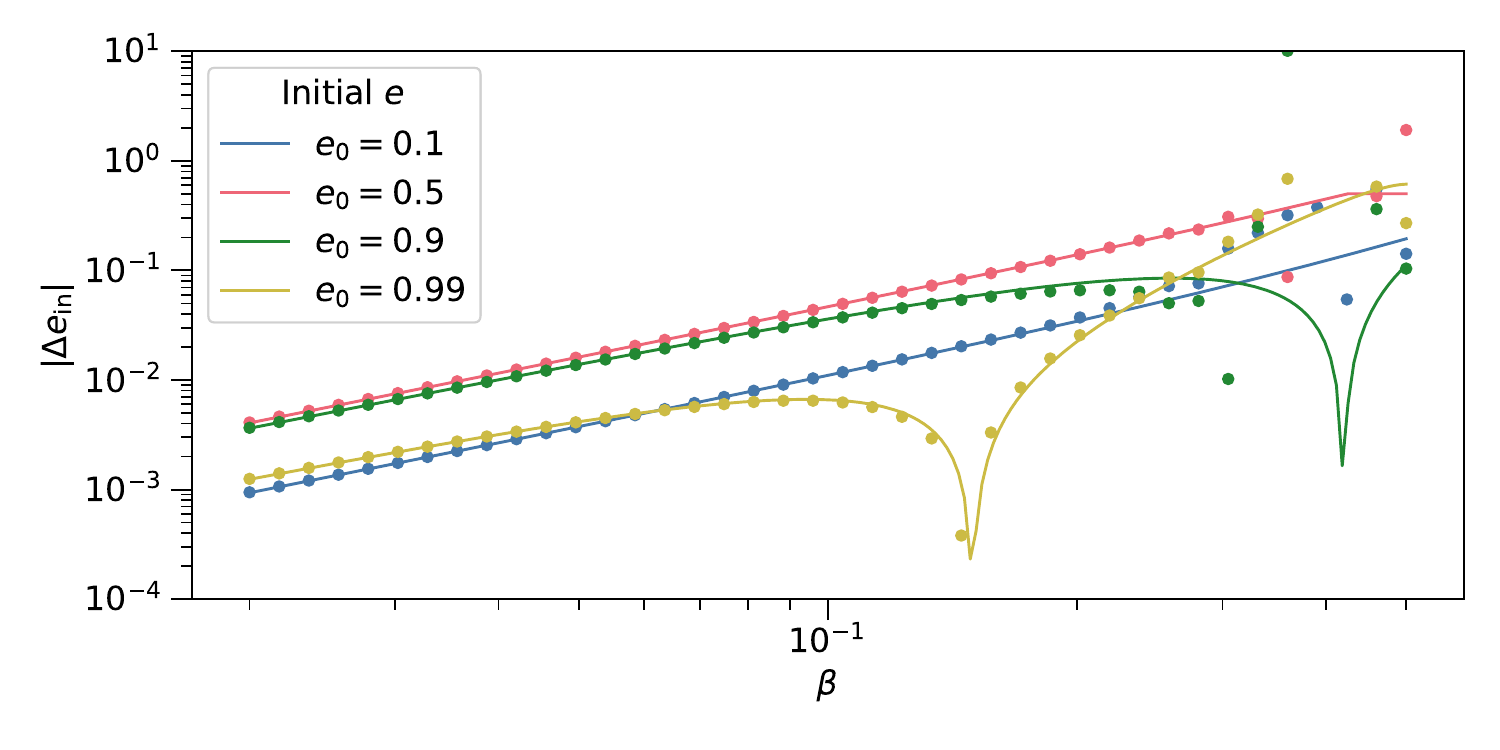}
    \caption{\label{fig:Revivals} Comparison of numerical (scatter points) and analytical (solid lines) scalar eccentricity changes. The numerical points were calculated using the REBOUND IAS15 integrator \citep{Rein2012, Rein2014}. The abscissa shows $\beta$ (log scale), and the ordinate shows the absolute change in eccentricity after one periapsis passage (log scale). The SMBH has mass $10^6\MSUN$, and the stars are solar-like with $\iota_{\mathrm{mut}}=1.2$ rad. Both analytical and numerical calculations assumed Newtonian three-body dynamics.}
\end{figure*}

\subsection{Hierarchical 3-body problem}
\label{sec:EOM}

Fig.~\ref{fig:sketch} sketches our set up and coordinate system. We adopt a Cartesian coordinate system centred on the SMBH, where the $\hat{\boldsymbol{x}}$-axis aligns with the outer orbit's semi-major axis, with apoapsis at positive $x$. The $\hat{\boldsymbol{z}}$-axis is aligned with the outer orbit's angular momentum vector. Because changes to the outer orbit are purely impulsive, we can update our coordinate system at each step to maintain this convention. Quantities corresponding to the inner and outer orbits are designated by lowercase and uppercase variables, respectively (e.g., $a_\inv$, $A_\outv$). The inner binary's argument of periapsis, $\omega$, longitude of ascending nodes, $\Omega$, and mutual inclination $\iota_{\rm mut}$, define the inner eccentricity vector as 

\begin{equation}
    \pmb{e}_\inv=e_\inv\begin{pmatrix} \cos\Omega\cos\omega-\sin\Omega\sin\omega\cos \iota_{\rm mut} \\ \sin\Omega\cos\omega+\cos\Omega\sin\omega\cos \iota_{\rm mut} \\ \sin \iota_{\rm mut} \sin\Omega \end{pmatrix},
\end{equation}
and the dimensionless angular momentum as
\begin{equation}
    \pmb{j}_\inv=\sqrt{1-e_\inv^2}\begin{pmatrix} \sin\Omega\sin \iota_{\rm mut} \\ -\cos\Omega\sin \iota_{\rm mut} \\ \cos \iota_{\rm mut} \end{pmatrix}.
\end{equation}
With $J_\outv$ aligned with the $z$-axis, the mutual inclination equals the inner orbit's inclination. These vectors always satisfy the orthogonality relations 
\begin{align}
    &e_\inv^2+j_\inv^2=1,\nonumber \\
    &\pmb{e}_\inv\cdot\pmb{j}_\inv = 0 \label{eq:orth}.
\end{align}

We always consider the system to be hierarchical, i.e. the inner semi-major axis is much smaller than the outer one $a_\inv/A_\outv\ll1$. Strong perturbations are expected for highly eccentric outer orbits, quantified by the pericenter $R_p=A_{\rm out}(1-E_{\rm out}) = R_{\rm TS}/\beta=a_{\rm in}(M_\bullet/m_{\rm in})^{1/3}/\beta$, where $E_{\rm out}$ is the outer orbital eccentricity. As long as $a_{\rm in}/A_{{\rm out}} \ll \beta (m_{\rm in}/M_\bullet)^{1/3}$, we can treat the encounter as parabolic and neglect the effects of the SMBH elsewhere. 

We use the approximation given in \cite{Hamers2019SecondOrder} to compute $\Delta \pmb{e}_{\text{in}}$ and $\Delta \pmb{j}_{\text{in}}$ during each periapsis passage. Expanding the three-body Hamiltonian to quadrupole order, and using second-order perturbation techniques, they obtain the changes in the inner orbit in their Eq. 24:
\begin{align}
    \Delta \boldsymbol{e}_{\rm in} = \varepsilon_{\rm SA} \boldsymbol{f}_e + \varepsilon_{\rm SA}^2 \boldsymbol{g}_e \nonumber \\
    \Delta \boldsymbol{j}_{\rm in} = \varepsilon_{\rm SA} \boldsymbol{f}_j + \varepsilon_{\rm SA}^2 \boldsymbol{g}_j,  \label{eq:de_dj} 
\end{align}
where the functions $\boldsymbol{f}_e, \boldsymbol{f}_j, \boldsymbol{g}_e$ and $\boldsymbol{g}_j$ are derived in \cite{Hamers2019SecondOrder} which we provide in Appendix \ref{appendix:hamers} for completeness. \epsSA \ is a small parameter given by $\varepsilon_{\rm SA}=(\beta/2)^{3/2}$. The calculation assumes that binary's mean angular velocity exceeds the SMBH's angular velocity at periapsis. This holds for $\beta \ll 1$ but can break down for larger $\beta$.

Fig.~\ref{fig:Revivals} shows the comparison between the analytic prediction and the numerical results. While the match is excellent for low $\beta$, it breaks down for $\beta \gtrsim 0.3$ where the evolution becomes chaotic. While the scalar change in eccentricity over a single outer orbit is reasonably well captured, small coherent errors may accumulate over multiple orbits. Accordingly, we adopt the conservative requirement that $\beta < 0.15$ throughout the remainder of this work.  Higher-order octupole terms from \cite{Hamers2019-cw} are negligible for our parameter regime. 

Under the secular approximation, $\Delta a_\inv=0$, and the evolution is governed solely by changes in $\pmb{e}_\inv$ and $\pmb{j}_\inv$. Over many orbital iterations,  the orthogonality relations in Eq.~\ref{eq:orth} are not preserved through repeated application of Eq. \ref{eq:de_dj} and drift with errors of order $\beta^3$. To enforce these constraints, we first use Eq.~\ref{eq:de_dj} to update the eccentricity ($\pmb{e}_\inv'=\pmb{e}_\inv+\Delta \pmb{e}_\inv$), and then reconstruct the new dimensionless angular momentum as follows.  We evolve $j_z \to j_z' = j_z + \Delta j_z$ where $\Delta j_z$ is taken from Eq.~\ref{eq:de_dj}, and then impose the orthogonality relations to find $j_x'$ and $j_y'$, yielding 

\begin{align}
    j_y'&=-\frac{e_y'e_z'j'_z\pm e_x'[-e_\inv'^2j_z'^2 + (e_x'^2+e_y'^2)(1-e_\inv'^2)]^{1/2}}{e_x'^2+e_y'^2}\nonumber \label{eq:2:jyPrime}.\\
    j_x'&=\frac{-e'_zj_z'-e_y'j_y'}{e_x'}.
\end{align}

\begin{figure*}[t!]
    \centering
    \includegraphics[width=0.98\textwidth]{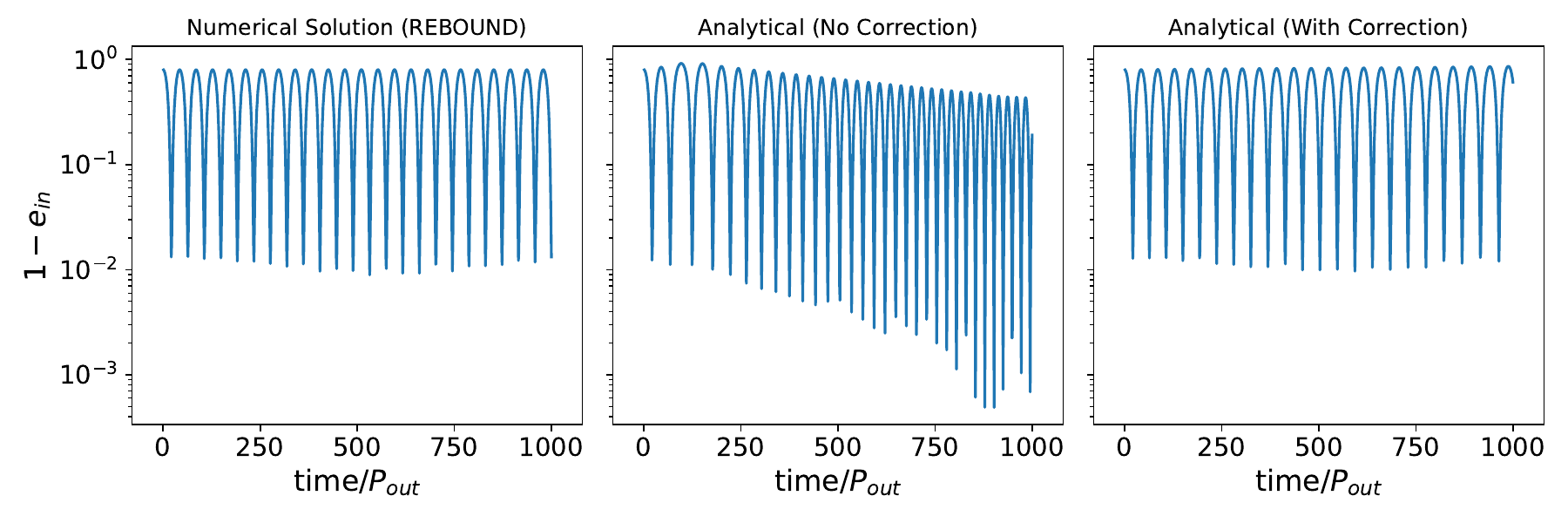}
    \caption{\label{fig:hammersCorrection} The evolution of the eccentricity of a typical binary using Newtonian physics only. On the left is a REBOUND numerical integration, in the middle is the semi-analytical calculation following  \citet{Hamers2019SecondOrder} without any corrections, while the calculation on the right employs our corrections. The oscillations above are ZLK oscillations that occur over tens of outer orbits. Note that the ordinate is logarithmic, and $r_p=a_\inv(1-e_\inv)$, so without corrections the binary stars artificially approach separations an order of magnitude closer than in reality.}
\end{figure*}

Eq.~\ref{eq:2:jyPrime} has two solutions, so we choose the solution which is closest to the uncorrected \cite{Hamers2019SecondOrder} expression. When the term in the square brackets is negative, we set it to $0$, leading to
\begin{equation}
    |j_z^{\text{final}}|=\sqrt{e_x'^2+e_y'^2-e^2_\inv(e_x'^2+e_y'^2)}/e_\inv'
\end{equation}
This corresponds to $j_z^{\text{final}}\approx (1\pm0.001)j_z'$. In $\sim1\%$ of cases, small coherent errors accumulate over thousands of outer orbits, but these are erased once two-body relaxation is included, as random angular momentum kicks disrupt the coherence.

The impact of this correction can be seen in Fig.~\ref{fig:hammersCorrection}, where we show the evolution of the binary's eccentricity over $1000$ outer orbits, computed with various methods including the one used in this work. Direct numerical integration using the full equations of motion (on the left) shows standard ZLK oscillations, while the approximate method without correction (middle plot) shows that the binary reaches spuriously high eccentricity. Using our method with the corrections outlined above, we see the standard ZLK oscillations again. 

\subsection{Stellar Tides} \label{sec:StellarTides}

Binaries at close separations and/or large eccentricities are affected by tidal evolution. We only consider tides raised by the stellar companion, since the tides raised by the SMBH are negligible.
In the weak-friction equilibrium tide model, a static tidal bulge is developed due to the response of a star to a tide raised by the companion  \citep{Zahn1977, Hut1981, Eggleton2001ApJ...562.1012E, Hurley2002, Belczynski2008}. The orbital evolution is given by
\begin{align}
    \deriv{e_\inv}{t}=-\frac{9e_\inv}{\tau_{\mathrm{TF},i}(1-e_\inv^2)^{13/2}} C_e(e_{\rm in}, \nu_i/\nu_{\rm in}), \\
     \deriv{a_\inv}{t}=-\frac{2a_\inv}{\tau_{\mathrm{TF},i}  (1-e_\inv^2)^{15/2}} C_a(e_{\rm in}, \nu_i/\nu_{\rm in}),
\end{align}
where the coefficients $C_e(e_{\rm in}, \nu_i/\nu_{\rm in}), C_a(e_{\rm in}, \nu_i/\nu_{\rm in})$ depend on the eccentricity $e_{\rm in}$ and on the ratio of the spin frequency $\nu_i$ and the orbital angular frequency $\nu_{\rm in}=(G m_{\rm in}/a_\inv^3)^{1/2}$.

Given that spin synchronisation occurs faster than orbital circularisation, we assume the binary rapidly reaches a pseudo-synchronous state where there is no further angular momentum exchange between the inner orbit and the stellar spins, and the stellar rotational frequencies $\nu_i$ are `almost' synchronised with the orbital period, and depend only on the orbital eccentricity \citep{Hut1981}:
\begin{equation}
    \frac{\nu_i}{\nu_\inv}=1+6e_\inv^2+\frac{3}{8}e_\inv^4+\frac{223}{8}e_\inv^6.
\end{equation}
This differs from \cite{Bradnick2017-ed} who assumed non-rotating stars ($\nu_i=0$), potentially overestimating tidal dissipation. This makes $C_e, C_a$ depend only on the orbital eccentricity (see \citealp{Hut1981} for full expressions and details).

The tidal friction timescale, $\tau_{\rm TF}$ can be estimated  \citep{Eggleton2001ApJ...562.1012E} as
\begin{equation}
    \tau_{\rm TF,i}= \frac{\tau_{v,i}}{9}\left(\frac{a_\inv}{r_i}\right)^8\frac{m_i^2}{m_\binv m_j}(1-Q_i)^2,
\end{equation}
where $m_j$ is the mass of the companion star, $r_i$ is the stellar radius of star $i$, and $Q_i$ is the quadrupole deformability of the star. $m_{\rm bin}=m_i+m_j$ is the total binary mass. We set $Q_i=0.021$, to be consistent with an $n=3$ polytrope star \citep{Eggleton1998}. $\tau_{v,i}$ are the viscous timescales and can be estimated by 
\begin{equation}
    \tau_{v,i}=100\left(\frac{3m_ir_i^2}{l_i}\right)^{1/3},
\end{equation}
where $l_i$'s are the stellar luminosities, which we set to be 
\begin{equation}
    \frac{l_i}{\LSUN}=\alpha\left(\frac{m_i}{\MSUN}\right)^\zeta, 
\end{equation}
where $\alpha$ and $\zeta$ are constants that depend on $m_i$ and are listed in Table \ref{fig:tableAlpha}. These fits are valid for ZAMS stars. 

\begin{table}
\label{fig:tableAlpha}
\centering
\begin{tabular}{ccc}
\hline
Mass ($\MSUN$) & $\alpha$ & $\zeta$ \\
\hline
$<0.43$ & 0.23 & 2.3 \\
0.43–2.00 & 1.00 & 4.0 \\
2.00–20.00 & 1.50 & 3.5 \\
$>20.00$ & 2700.00 & 1.0 \\
\hline
\end{tabular}
\caption{Parameters used in the mass–luminosity relationship \citep{salaris2005, Duric2012-qv}.} 
\label{tab:masslum}
\end{table}

\subsubsection{Dynamical tides for highly eccentric binaries}

For highly eccentric orbits the weak friction limit breaks down and the perturbation becomes impulsive, leading to simultaneous excitations of multiple stellar oscillations modes \citep{Fabian1975-xs, Press1977-dt}. We use a simplified implementation presented by \cite{MoeKratter2018}. The energy deposited in star $i$ is 
\begin{align}
    \Delta \mathcal{E}_i=f_{i,\mathrm{dyn}}\frac{m_\binv}{m_i}\frac{Gm_2^2}{r_i}\left(\frac{r_{p,\inv}}{r_i}\right)^{-9},
\end{align}
where $f_{\mathrm{dyn}}$ is a dimensionless efficiency factor. As dynamical tides are only applied to highly eccentric binaries, this energy loss occurs at periastron, leaving $r_{p,\inv}$ constant, and shrinks the binary according to, 
\begin{align}
    \deriv{a_\inv}{t}=-\frac{2 a_\inv^2}{Gm_1m_2 P_\inv}(\Delta \mathcal{E}_1 + \Delta \mathcal{E}_2),
\end{align}
where $P_\inv$ is the orbital period of the inner binary. 


To complete this prescription, we need to provide a value for the transitional eccentricity $e_c$ where dynamical tides are turned off and for the efficiency factor $f_{\rm dyn}$. However, these are heavily debated in the literature. The conditions for dynamical tides are largely independent of the semimajor axis for a wide enough binary and depend mostly on the pericenter \citep{gp22, dodici25}. \cite{MoeKratter2018} used $e_c=0.8$ to be compatible with the transition from the chaotic boundary to regular tidal evolution \citep{mardling1, Mardling1995-vu}. \cite{marklund2025} also used a similar approach for the transition eccentricity based on \cite{Mardling1995-vu}. \cite{dodici25} assumed a far more efficient mechanism where chaotic tides efficiently operate until the binary is effectively circularised. \cite{gp22} constructed a continuous transition between equilibrium and dynamical tides. Since we are primarily interested in tight binaries, $a_\inv \le 1\ \rm au$, \cite{MoeKratter2018}'s prescription is relevant, and we follow them and set $e_c=0.8$. 

The efficiency factor can vary by an order of magnitude. \cite{MoeKratter2018} used $f_{\rm dyn}=0.3$, however, there are arguments that $f_{\rm dyn}$ could be smaller by a factor $10$ \citep{McMillan1986-uq, Mardling1995-vu, Lai1997-ix}. We set $f_{\rm dyn}=0.3$ following \cite{MoeKratter2018}. We've tried other values and found that the results are mostly independent of $f_{\rm dyn}$.

\subsection{Apsidal Precession}
\label{sec:precession}
GR precession is included via the 1PN approximation, where the change in the argument of pericenter $\Delta \omega$ per orbit is
\begin{equation}
    \label{eqn:GRPrecessionSingleOrbit}
    \Delta \omega = \frac{6\pi G m_{\rm bin}}{c^2 a(1-e^2)},
\end{equation}
where $c$ is the speed of light. The precession timescale $2\pi P / \Delta \omega$ is negligible for the outer orbit ($\sim10^8$ yr). However, the inner orbit can be significantly altered; in particular, when the 1PN precession timescale is comparable to the ZLK oscillation timescale, ZLK oscillations can be quenched. For a direct comparison, we rewrite 
\begin{equation}
    \left(\deriv{\omega}{t}\right)_{\mathrm{GR}}=\frac{\varepsilon_{\mathrm{GR}}}{\tau_{\mathrm{ZLK}}(1-e_\inv^2)},
\end{equation}
where
\begin{equation}
    \varepsilon_{\mathrm{GR}} = \frac{3Gm_\binv^2A_\outv^3(1-E_\outv^2)^{3/2}}{a_\inv^4 c^2 \BHM}
\end{equation}
\citep{Liu2015-qt, mangipudi22}, and
 $\tau_{\mathrm{ZLK}}$ is the characteristic ZLK timescale, given by \citep{Liu2015-qt}\footnote{This is an approximate timescale, while the actual libration or circulation timescale depends on the initial conditions of the triple system \citep{kn2007, antognini15, grishin24}.} as
\begin{equation}
    \tau_{\mathrm{ZLK}} \approx \frac{1}{2\pi}\frac{P_\outv^2}{P_\inv}(1-E_\outv^2)^{3/2}\approx0.5\beta^{-3/2}P_\outv.
    \label{eq:tauZLK}
\end{equation}
Similarly, the rate of precession from equilibrium tides is 
\begin{equation}
    \left(\deriv{\omega}{t}\right)_{\mathrm{tide},i}=\frac{\varepsilon_{\mathrm{tide},i}}{\tau_{\mathrm{ZLK}}(1-E_\outv^2)^5}\left(1+\frac{3}{2}e_\inv^2+\frac{1}{8}e_\inv^4\right),
\end{equation}
with 
\begin{equation}
    \varepsilon_{\mathrm{tide},i}=\frac{15m_jm_\binv A_\outv^3(1-E_\outv^2)^{3/2}k_ir_i^5}{a_\inv^8m_i\BHM},
\end{equation}
where $k_i$ is the Love number, which can be related to the quadrupole deformability $Q_i$ by $k_i = Q_i/(1-Q_i)$  \citep{Eggleton2001ApJ...562.1012E}.

Finally, the total extra precession due to tides and GR is
\begin{equation}
    \deriv{\omega}{t}=\left(\deriv{\omega}{t}\right)_{\mathrm{tide},1}+\left(\deriv{\omega}{t}\right)_{\mathrm{tide},2}+\left(\deriv{\omega}{t}\right)_{\mathrm{GR}}.
\end{equation}

\subsection{Two-Body Relaxation}

\begin{figure*}[]
    \centering
    \includegraphics[width=0.85\textwidth]{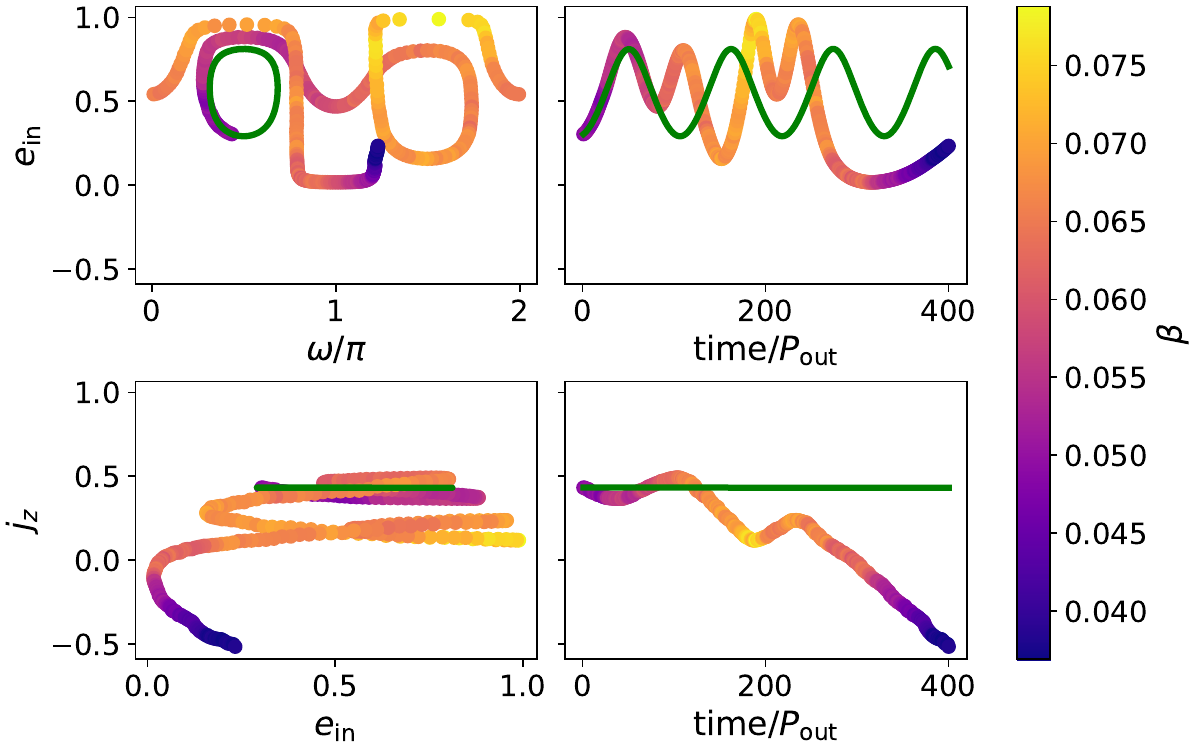}
    \caption{\label{fig:singleKozaiCombined} The evolution of the inner orbit of a typical stellar binary, when including no kicks (green), and kicks (purple and orange scatter points). There is no additional physics considered here. The colour represents $\beta$, dark purple being lower (further from the SMBH), and orange being higher (closer).}
\end{figure*}

Encounters with background stars perturb the outer orbit's energy and angular momentum. The rate of encounters with impact parameter $b$ at distance $R$ from the SMBH is 
\begin{equation}
    \label{eqn:EncounterRate}
    \Gamma(b,R)=\pi n(R)b^2\sigma(R),
\end{equation}
where $n(R)$ is the local stellar number density, $\sigma(R)=\sqrt{G\BHM/R}$ is the typical velocity dispersion within the sphere of influence at distance $R$ from the SMBH and we have assumed $\sigma(R) \gg \sqrt{G m_* / a_{\rm in}}$, so gravitational focusing can be ignored. Over one outer orbit, this gives a change in specific angular momentum of (see Section \ref{sec:two-bodyAppendix})
\begin{align}
\label{method:eqn:DeltaJ}
    \langle \Delta J_\outv^2 \rangle \approx1.3\frac{J_c^2(R_{a,\outv})}{\tau_{\mathrm{relax}}(R_{a,\outv})}P_\outv,
\end{align}
where $J_c(R_{a,\outv})=\sqrt{G\BHM R_{a,\outv}}$ is the specific angular momentum of a circular orbit around the SMBH with apoapsis $R_{a,\outv}$, which is given by $R_{a,\outv}=A_\outv(1+E_\outv)\approx 2A_\outv$. $\tau_\text{relax}(A_\outv)$ is the two-body relaxation time \citep[e.g.,][]{Spitzer1971-bu}:
\begin{align}
    \tau_\text{relax}=\frac{\sigma^3}{15.4G^2nm_*^2 \ln \Lambda},
\end{align}
where $m_*=0.5\MSUN$ is the typical stellar mass, and $\ln \Lambda=\ln (0.4N)$ is the Coulomb logarithm, where $N = \BHM / m_*$ is the number of stars in the sphere of influence.

Performing orbit-averaging for the $x$, $y$ and $z$ components of angular momentum, we find that 
\begin{equation}
    \langle \Delta J_{\outv,x}^2 \rangle\approx 0.86(1-E_\outv) \langle \Delta J_{\outv,y}^2 \rangle,
\end{equation}
and $ \langle \Delta J_{\outv,y}^2 \rangle \approx \langle \Delta J_{\outv,z}^2 \rangle$. The full details of the orbital averaging procedure the calculation of individual orbit-averaged quantities enlisted in Appendix \ref{appendix:A}. For an eccentric orbit, angular momentum relaxation is dominated by interactions near apoapsis, which produce little change in the angular momentum component along the $x$ axis. Thus, unlike the isotropic 3D anguler momentum kicks in \cite{Bradnick2017-ed}, we consider only 2D angular momentum kicks in the $y-z$ plane. 

We additionally consider changes in the specific energy of the outer orbit, $\mathcal{E}_\outv(A_\outv)= - G\BHM/2A_\outv$, due to two-body relaxation. Each flyby changes the specific energy of the outer orbit by $\Delta \mathcal{E}_\outv\approx (Gm_*/b\sigma)^2$. Integrating over an entire orbit using the framework in Section~\ref{sec:two-bodyAppendix} yields the mean-square energy change per orbit:

\begin{equation}
\label{Methods:eqn:deltaE}
     \langle \Delta \mathcal{E}_\outv^2 \rangle\approx0.87\frac{\mathcal{E}^2(R_{p})}{\tau_{\mathrm{relax}}(R_{p})}P(R_p),
\end{equation}
where $P(R_p)$ and $\mathcal{E}(R_p)$ are the period and energy of a circular orbit with semi-major axis $R_p$. For a Bahcall-Wolf cusp density profile (see sec. \ref{sec:MonteCarlo}), energy changes are dominated by close passages at periapsis, while angular momentum changes are dominated by encounters at apoapsis \citep{SariFragione2019}. Therefore, for each outer orbit, we apply an impulsive energy kick at periapsis drawn from the normal distribution $\mathcal{N}(0, \langle \Delta \mathcal{E}_\outv^2 \rangle)$, followed by an angular momentum kick at apoapsis whose $y$ and $z$ components are drawn independently from $\mathcal{N}(0, \langle \Delta J_\outv^2 \rangle/2)$.

\section{Analysis of a Single System} \label{sec:singleSysResults}

\subsection{Angular Momentum Kicks and ZLK Oscillations}


To illustrate the key dynamical processes, we first examine the evolution of individual binaries. The simplest case is Newtonian gravity without any perturbations, where a system exhibits ZLK oscillations between $e_{\inv,\min}$ and $e_{\inv, \max}$ with fixed $j_z$ (green lines in Fig.~\ref{fig:singleKozaiCombined}). Including angular momentum kicks produces qualitatively different, stochastic behaviour (Fig.~\ref{fig:singleKozaiCombined}). The $j_z$ is no longer conserved. Despite this, the system still traces ZLK-like tracks, where the evolution of $\omega$ changes back and forth from libration around $\pi/2, 3\pi/2$ to full circulations. The oscillations occur on timescales comparable to $\tau_{\rm{ZLK}}\sim50P_\outv$, confirming that the dynamics remain broadly ZLK-like (cf.~Eq.~\ref{eq:tauZLK}). These kicks can drive binaries to even higher eccentricities than standard ZLK oscillations, increasing the likelihood of merger. 


This stochasticity arises because angular momentum kicks over a full orbital period collectively rotate the outer orbital plane about the $x$-axis by an angle $\psi \approx \langle|\Delta J_\outv|\rangle / J_\outv$. As we are considering binaries in the empty loss-cone, we have $\psi\ll 1$. The resulting change in the inner angular momentum vector in the reference frame fixed to the outer binary is 
\begin{align}
   \label{fig:djAngleKick}
   \Delta \pmb{j}_\inv = \left[0, \hspace{6pt}\frac{\psi^2}{2}j_y+\psi j_z,\hspace{6pt}\frac{\psi^2}{2}j_z-\psi j_y  \right].
\end{align}
When the binary reaches the extrema of its eccentricity oscillations ($e_{\inv,\max}$ or $e_{\inv, \min}$), ZLK evolution changes $j_\inv$ slowly, allowing these random kicks to dominate. This is evident in the left panels of Fig.~\ref{fig:singleKozaiCombined}, where the binary switches between ZLK-like tracks near the turning points. 

As shown in the bottom left panel of Fig.~\ref{fig:singleKozaiCombined}, $j_z$ is no longer conserved. The largest deviations occur near $e_{\inv, \min}$ where $j_y$ is largest, consistent with Eq.\eqref{fig:djAngleKick}. 

The angular momentum kicks also induce a random walk in the outer periapsis distance, and hence in $\beta$. The diffusion timescale between two factors $\beta_{\mathrm{initial}}$ and $\beta_{\mathrm{final}}$ is 
\begin{equation}
    \tau_{\mathrm{relax}}(\beta_{\mathrm{initial}}, \beta_{\mathrm{final}})=\frac{J_{\mathrm{LC}}^2\left(\beta_{\mathrm{initial}}^{-0.5}-\beta^{-0.5}_{\mathrm{final}}\right)^2}{\langle \Delta J_\outv^2 \rangle}P_\outv .
\end{equation}
For our systems, transitioning from $\beta=0.05$ to $\beta=0.15$ we find that $\tau_{\mathrm{relax}}(0.05,0.15)\sim(10^3-10^4)P_\outv$. This is too slow to affect the ZLK-like oscillations discussed above. We'll distinguish between pure ZLK oscillations (without perturbations) and perturbed ZLK oscillations that vary $j_z$ and the eccentricity over longer timescales. 

\subsection{Effects of General Relativity and Tides}
GR precession acts to shorten the ZLK timescale and suppress oscillations. Since $\varepsilon_{\mathrm{GR}} \propto a_\inv^{-4}$, this effect is strongest for tight binaries, effectively shielding them from tidally induced mergers. Stellar tides can enhance this suppression by circularising the orbits before high eccentricities are reached. We find that when $\varepsilon_{\rm GR} > 7$, GR precession dominates and no significant ZLK oscillations occur, whereas for $\varepsilon_{\rm GR}\ll 1$, GR is negligible. This is explored further in Section \ref{sec:MonteCarlo}. 

\subsection{Stellar Encounters}\label{sec:SE}
So far we have neglected the impact of other stars on the inner binary's orbital parameters. These encounters can change both the energy and angular momentum of the inner binary. 

When considering energy changes due to stellar interactions, we may split our systems into two regimes, hard and soft. When the magnitude of the binary's orbital energy $G m_1 m_2 /(2a_\inv)$ exceeds the kinetic energy of a typical star $m_*\sigma^2/2$, the binary is hard;  otherwise it is soft \citep{Heggie1975}. The hard-soft boundary at distance $R$ from the SMBH is given by 
\begin{equation} \label{eq:a_sh}
    a_{\rm in, hs}(R) = 0.2 \left( \frac{R}{\rm pc}\right) \left(\frac{ m_*}{ M_\odot}\right)^{-1}\left(\frac{ m_1m_2}{M_\odot^2}\right) \left( \frac{\BHM}{10^6 M_\odot}\right)^{-1}\ \rm au\ .
\end{equation}
For our systems, this boundary is at $a_{\inv, hs}\approx0.1$au. This approximately matches the lower limit of the binary periods, $\sim 10$ days, that we consider (see Section \ref{sec:MonteCarlo}), hence we expect most binaries to be soft. 

Hard binaries will tend to become harder, and soft binaries will tend to become softer. The timescale for hardening an already hard binary is given by \citep{Heggie1975, HeggieHut1993} 
\begin{equation}
    \tau_{\mathrm{hard}}\sim 10 \left(\frac{\sigma}{65\ \rm km/s}\right) \left(\frac{\rho}{10^5\MSUN {\rm pc}^{-3}}\right)^{-1}  \left(\frac{a_\inv}{0.8\ \rm au}\right)^{-1}\ \rm Gyr,
\end{equation}
where $\rho$ is the local mass density of stars at apoapsis. We find for systems on the hard-soft boundary, $\tau_{\mathrm{hard}}\sim 10^{10}$yr, much longer than our simulation durations ($10^{7}$ to $10^8$yr). Thus, hardening can safely be neglected (although binaries can be hardened by tides on faster timescales, \citealp{dodici25}). 

Soft binaries evaporate on a timescale
\begin{equation}
    \tau_{\mathrm{evap}}\sim 1 \left(\frac{\sigma}{65\ \rm km/s}\right) \left(\frac{\rho}{10^5\MSUN \rm pc^{-3}}\right)^{-1}\left(\frac{\ln \Lambda}{3}\right)^{-1}  \left(\frac{a_\inv}{0.6\ \rm au}\right)^{-1}\ \rm Gyr,
\end{equation}
which is typically few times $\sim 10^9$ years near the hard-soft boundary, but can fall to $\sim 10^8$yr for few $a_\inv\sim1$AU. Evaporation, or at least scatterings, may therefore affect the widest systems (see \citealt{hopman09} and the derivations in the appendix).

\section{Monte Carlo Simulation}
\label{sec:MonteCarlo}

\begin{figure}[]
    \centering
    \includegraphics[width=0.4\textwidth]{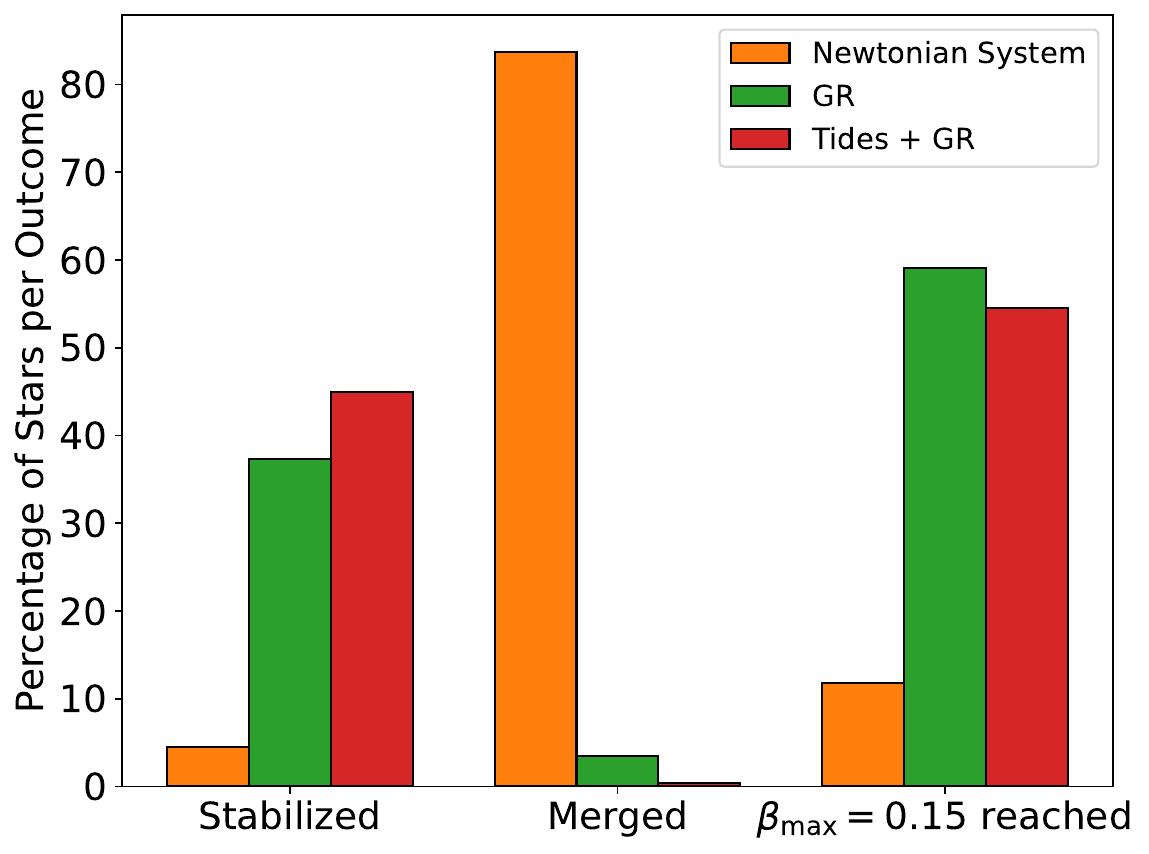}
    \caption{\label{fig:PercentagePLot} Bar graph of outcomes for stellar binaries. There is a negligible fraction of surviving binaries which are not shown here. There are 3 simulation sets: Newtonian (orange), GR included (green) and tidal forces and GR included (red).  The percentage of systems with a particular outcome is shown per simulation set.}
\end{figure}

\subsection{Initial Conditions} 
\label{sec:InitialConditions}
Here we describe the initial conditions of our simulations, which are primarily sampled from the empirical distributions of \citet[][hereafter MD17]{MoeStefano2017}, unless stated otherwise.

The primary mass $m_1$ is drawn from the Kroupa initial mass function \citep{Kroupa2001MNRAS.322..231K}, within $m_1\in[0.1,15]\MSUN$. The secondary mass follows the MD17 mass ratio distribution $q=m_2/m_1$. The radius of each star is given by $r_i=(m_i/\MSUN)^k\RSUN$, where $k=0.8$ if $m_i<\MSUN$, and $k=0.6$ otherwise \citep{Kippenhahn1994sse..book.....K}. 

The period, $P_\inv(m_1)$ is drawn from the MD17 distribution. Most common low-mass systems follow \"Opik's law \citep{Opik1924PTarO..25f...1O}, where $p(P_\inv)\propto1/P_\inv$. 

Binary separations and periods are restricted to $a_\inv\leq 1$ au, and $P_\inv\in[10 \text{ days}, 1 \text{ year}]$, respectively. Tight $P_\inv\lesssim10$ d binaries are essentially unchanged. According to Eq. \ref{eq:a_sh} and Kepler's law, binaries with typical masses $m_1=m_2=m_* = 0.5 M_\odot$ will be soft for $P \gtrsim 11.6\ (R/\rm pc)^{3/2}\ \rm days$, thus most binaries with $P_\inv \gtrsim 10$ days are expected to be soft.

The eccentricity is drawn from $p(e_\inv)\propto e_\inv^{\eta}$, where $\eta$ is given in MD17. We restrict the eccentricity so that the binary has an initial periapsis $r_{p,\inv}>1.2r_{p,\mathrm{crit}}$, where $r_{p,\mathrm{crit}}$ is the critical separation at which one of their stars will fill its Roche Lobe \citep{Eggleton1983ApJ...268..368E}. We also ensure $e<0.8$, so that the binaries do not start in the dynamical tidal regime. 

The inner and outer argument of periapsis, $\omega$, and longitude of the ascending node, $\Omega$, are uniformly drawn from $[0,2\pi]$, and the mutual inclination, $\theta = \cos \iota_{\rm mut}$, where $\iota_{\rm mut}=0$ for aligned orbital planes, is drawn uniformly from $[-1,1]$.

Following \cite{Bradnick2017-ed}, we consider binaries to be in a Bahcall-Wolf cusp \citep{Bahcall1976-bk} extending to 1 pc, with  $N=10^6$ stars, and number density $n(R)\propto R^{-7/4}$. The central SMBH has mass $\BHM=10^6\MSUN$.

The outer orbit's periapsis is initially determined by $\beta=0.05$, with an apoapsis distance drawn from the Bahcall-Wolf cusp within the range $R_a\in[100\ R_p, 1\ \rm{pc}]$, yielding a minimum outer eccentricity $E_\outv\geq0.98$.

\subsection{Stopping conditions}
 We simulate 100,000 binaries in three different cases: (1) no additional physics (we refer to this as the `Newtonian Model'), (2) with GR precession included, and (3) with tidal forces and GR included, resulting in 300,000 systems in total. Each simulation is run until one of the following four conditions is met:
\begin{enumerate}
    \item $\bMax=0.15$ reached: \textit{When $\beta\geq0.15$,  the analytical approximation begins to become unreliable.}
    \item Stabilized: \textit{When $\beta$ is sufficiently small, the perturbations to the inner binary from SMBH tidal forces become negligible, and the evolution is instead dominated by interactions with field stars. In our simulations, we take $\beta < 0.005$ as the criterion for stabilization. Additionally, when $E_{\rm out} \leq 0.97$, the parabolic approximation is no longer valid. We find that $\sim85\%$ of systems satisfying $E_{\rm out} \leq 0.97$ also have $\beta < 0.01$, and therefore we combine these two conditions.}
    \item Merged: \textit{We define a merger to occur when one of the stars has overfilled its Roche Lobe \citep{Eggleton1983ApJ...268..368E}. }
    \item Survived: \textit{If the binary survives for more than 200,000 outer orbits (roughly $10^9$ years).}
\end{enumerate}

\begin{figure}[]
    \centering
    \includegraphics[width=0.4\textwidth]{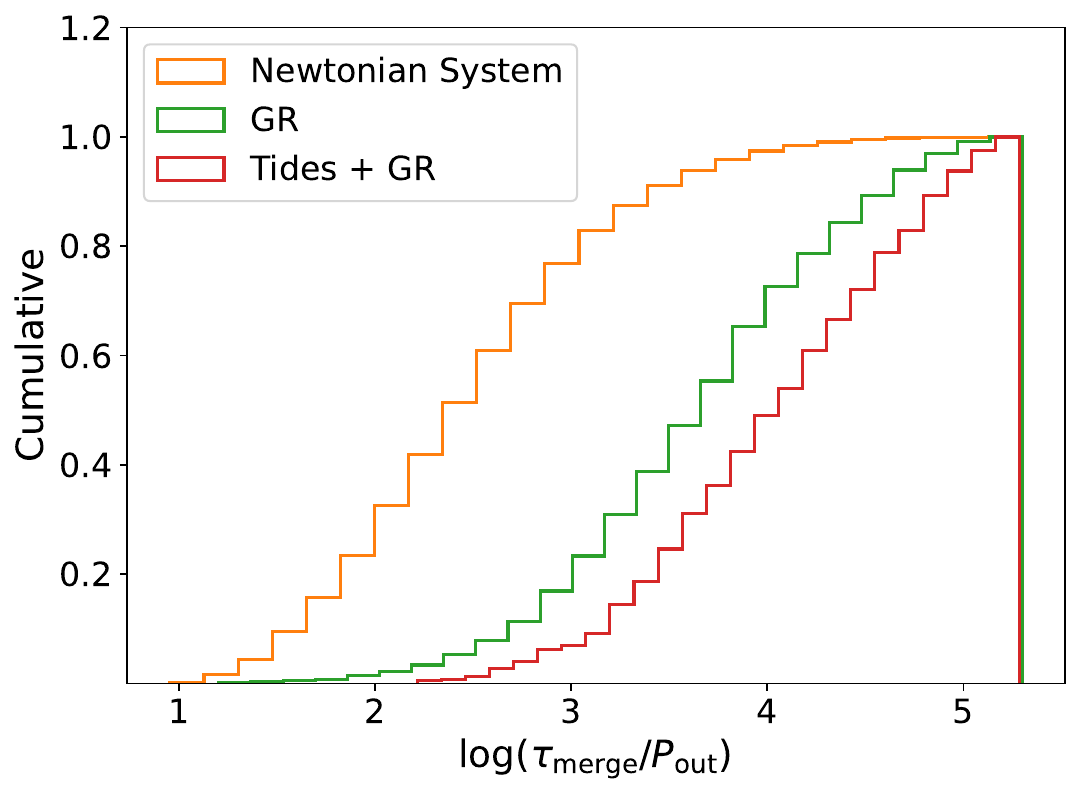}
    \caption{\label{fig:tcollide} Normalised cumulative distribution (CDF) of the merger time. The abscissa is log(number of outer orbits until merger), and the ordinate is the CDF. The orange, green and red lines represent the Newtonian models, systems with GR only, and systems with GR and tides included, respectively. }
\end{figure}

\begin{figure}[]
    \centering
    \includegraphics[width=0.4\textwidth]{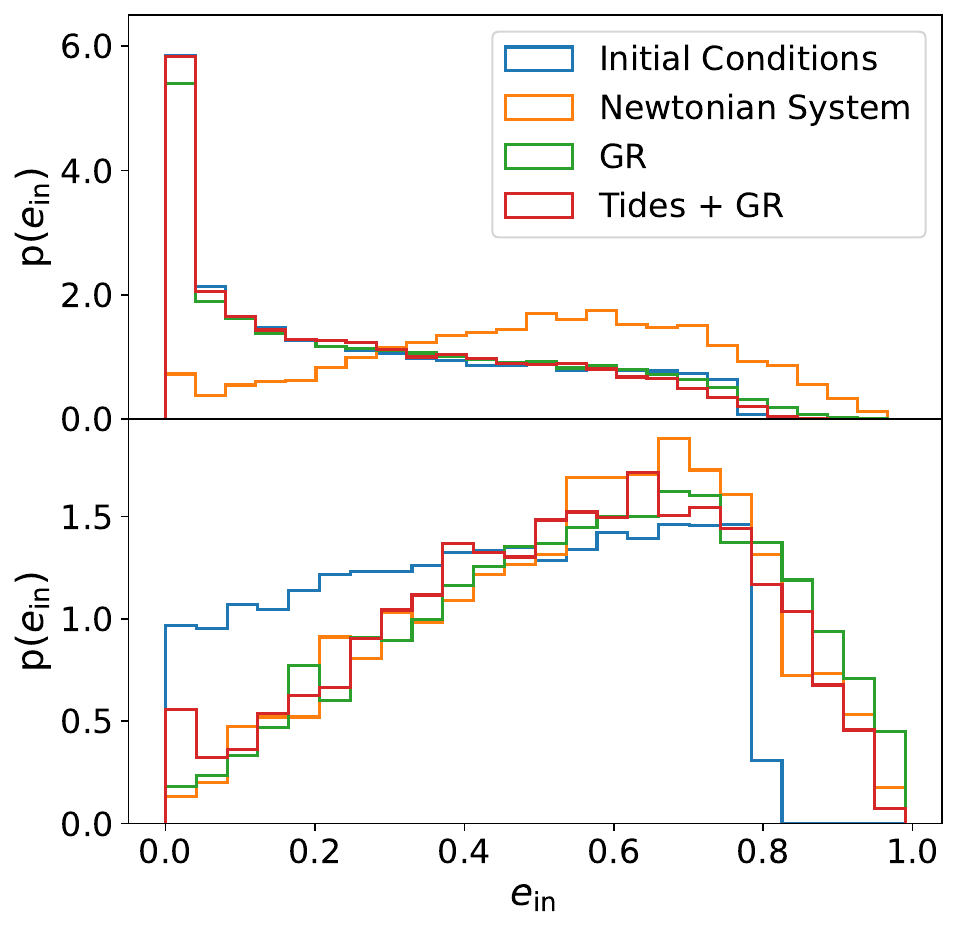}
    \caption{\label{fig:EDistribution} Normalised eccentricity distribution $e_\inv$ for systems that reached \bMax$=0.15$. The top panel shows systems with initial $\epsGR>7$, while the bottom panel shows systems with initial $\epsGR<1$. The blue-line is the initial eccentricity distribution. The orange, green and red lines represent the Newtonian models, GR only, and GR and tides, respectively. Note that $\epsGR$ is still defined for the Newtonian model for comparison.} 
\end{figure}

\subsection{Outcome Frequencies}

\begin{figure*}[t!]
    \centering
    \includegraphics[width=0.75\textwidth]{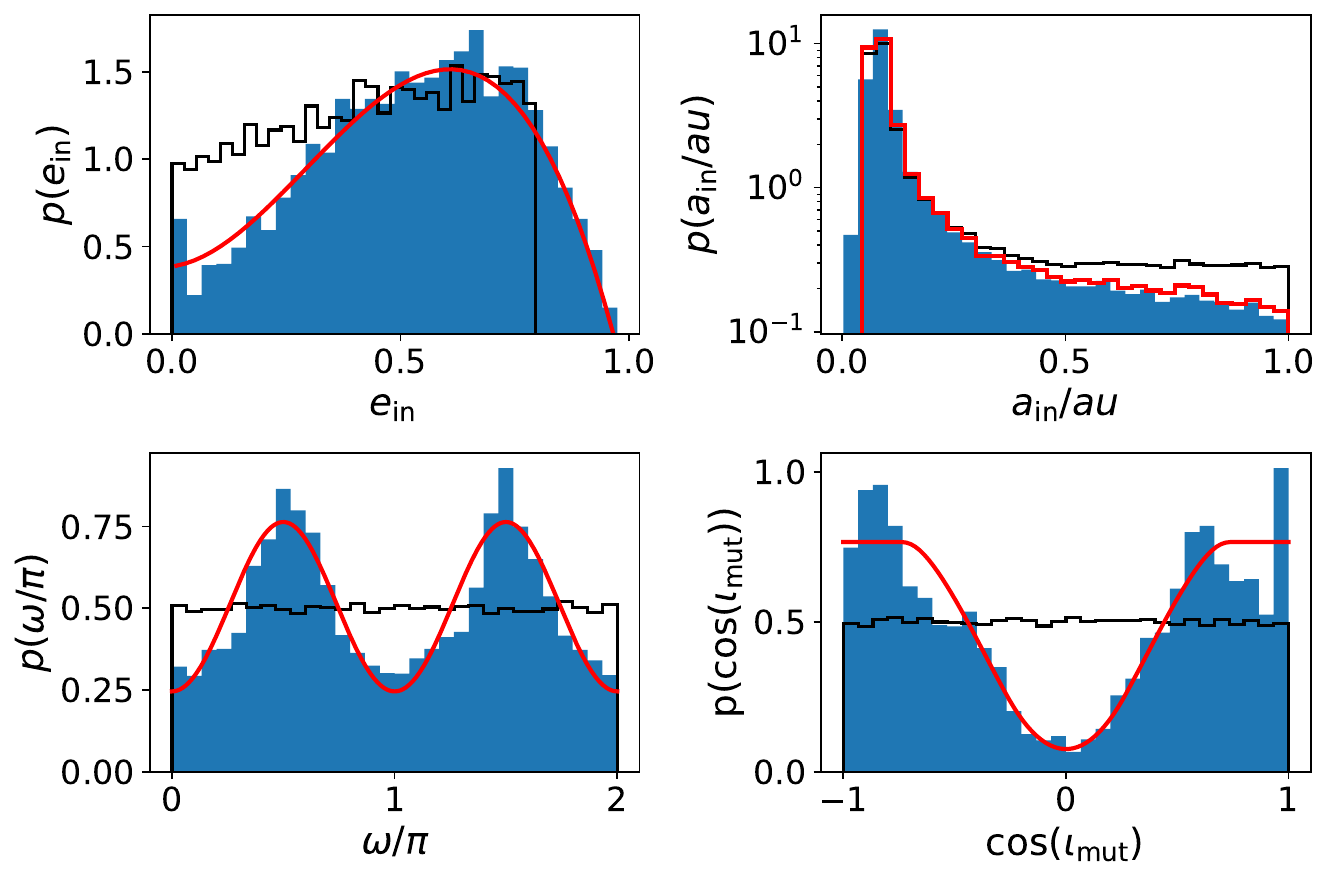}
    \caption{\label{fig:fitPlots} Fitted distributions for the binary parameters of systems when they reach $\bMax$. The simulations were done with kicks, GR and tides included. The filled blue histogram is the simulated distribution, the black lines are the initial distribution and  the red lines are the fitted distributions. Top left: eccentricity $e_\inv$. Top right: semi-major axis $a_\inv$. Bottom left: argument of periapsis $\omega$. Bottom right: inclination $\theta=\cos\iota_{\rm mut}$. All except the top right panel are for systems with $\epsGR<1$ for illustrative purposes. The top right panel has a logarithmic ordinate.}
\end{figure*}

We find that in the Newtonian model, ZLK oscillations increase inner binary eccentricities and drive the majority of binaries to merger.  Even in pure three-body Newtonian dynamics without angular momentum kicks, 63\% of binaries merge.  Momentum kicks replenish the parameter space of highly inclined orbits, which leads to larger eccentricities \citep[e.g., ][]{hamers_baror18} and an enhanced fraction of mergers: 84\%, as shown in Fig.~\ref{fig:PercentagePLot}. As shown in Fig.~\ref{fig:tcollide}, most mergers occur within the first $10^2-10^3$ outer orbits, which is comparable to the ZLK timescale. A smaller fraction merge at later times, driven by gradual inclination diffusion from successive angular momentum kicks. Due to the high merger rate, we find only $\sim 12\%$ of Newtonian systems reach $\beta_{\mathrm{max}}$. The remaining $\sim 4\%$ of stabilized systems are mostly wider binaries. This naturally follows from $E_\outv=1-\beta^{-1}A_\outv^{-1} a_\inv(\BHM/m_\inv)^{1/3}$, i.e., for larger $a_\inv$, the outer eccentricity is smaller for a fixed $A_\outv$, making it easier for binaries to leave the domain before reaching high $\beta$.

Including GR dramatically reduces the merger rate to $\sim3\%$, allowing $\sim 59\%$ of systems to reach \bMax.  GR precession can efficiently suppress ZLK oscillations, thus preventing mergers and allowing binaries to approach closer to the SMBH. Of the systems that merge, we find that almost all have $\epsGR < 1$. Adding tides increases the stabilized system fraction to $\sim 46\%$, at the expense of the mergers, which falls 10-fold down to $0.4\%$. This has little effect on the systems that reach \bMax\ since this outcome depends on relaxation processes near apocentre and not on the detailed evolution of the inner binary.

As illustrated in Fig.~\ref{fig:tcollide}, the timescales for binary mergers are longer when relativistic precession and tidal dissipation are included.  The merging systems are likely to be wider binaries with lower $\epsGR$ that are gradually perturbed into highly inclined orbits, where the ZKL mechanism allows them to reach high eccentricities, while relativistic and tidal effects suppress extreme eccentricity growth across the bulk of the population.

\subsection{Eccentricity Distributions}
Fig.~\ref{fig:EDistribution} shows the final eccentricity  distribution for binaries once they reach \bMax$=0.15$. For strong GR (initial $\epsGR > 7$, top panel), ZLK oscillations are effectively suppressed by GR and the final distribution is similar to the initial one. Small diffusion toward higher eccentricities still occurs, with only $\sim3.5\%$ obtaining eccentricities higher than $0.8$ in the GR only case. Adding tidal forces effectively quenches this possibility. For the weak GR case, (initial $\epsGR < 1$, bottom panel) the final eccentricity distribution is largely unaffected when compared to the Newtonian case (although GR decreases the number of mergers). Including tides shifts the distribution slightly toward lower eccentricities.

The probability distribution for systems with $\epsGR<1$ can be fit with (see the top left panel of Fig.~\ref{fig:fitPlots}): 
\begin{equation}
    \label{eqn:eDistProb}
    p(e_\inv)=0.39+0.46e_\inv+7.12e_\inv^2-7.96e_\inv^3.
\end{equation}

In the intermediate regime, ($\epsGR\in[1,7]$, about $32\%$ of systems) it is harder to describe the distribution. We suggest interpolating between Eq.~\eqref{eqn:eDistProb} and the initial eccentricity distribution. 

\subsection{Semi-major Axis Distribution}
The final semi-major axis distribution (Fig.~\ref{fig:fitPlots}) is best described through the conditional probability $p(\bMax|a_\inv)$, which is the probability of a particular system reaching \bMax\ given an initial $a_\inv$. In the Newtonian case, $p(\bMax|a_\inv)$ is largely independent of $a_\inv$. When GR is included, this probability declines approximately linearly with increasing $a_\inv$, consistent with the scaling of the GR precession rate (Eq.~\eqref{eqn:GRPrecessionSingleOrbit}). We find a fit given by: 
\begin{equation}
    p(\bMax|a_\inv)=\max\left[0.6-0.35\left( \frac{a_\inv}{\rm au}\right),0\right].
\end{equation}
Including tides produces negligible changes. The red line in the top right panel of Fig.~\ref{fig:fitPlots} shows this probability distribution based on the initial distribution of $a_\inv$. 

The latter probability fit does not include the possibility that a binary is unbound by field stars. Stellar interactions in the galactic centre (Section \ref{sec:SE}) can unbind the widest binaries. This would likely cause there to be fewer wide binaries reaching \bMax.  


\subsection{Inclination Distribution} 
The bottom left panel of Fig.~\ref{fig:fitPlots} shows the inclination distribution of our simulated systems that reached \bMax for the weak GR case ($\epsGR<1$). We can see that there is a depletion of systems at high inclinations, where ZLK oscillations are strong. For systems with strong GR ($\epsGR>7$), we find the distribution becomes flat.    

When GR is included, assuming the initial eccentricity $e_0\approx 0$, the extent of the maximal eccentricity is somewhat limited by \citep{Liu2015-qt}
\begin{equation}
    j_{\inv, \mathrm{ min}}=\frac{1}{9}\left[ 4\epsGR + \sqrt{16\epsGR^2+135\theta_0^2}\right].
\end{equation}
This gives a maximum eccentricity of $e_{\mathrm{max}}^2=1-j_{\inv, \mathrm{ min}}^2$, where $\epsGR<1$. Setting $e_{\mathrm{max}}^2=0$ defines the critical ZLK inclinations,
\begin{equation}
    \label{eqn:thetac}
    \theta_c(\epsGR)=\pm\sqrt{\frac{1}{15}(9-8\epsGR)},
\end{equation}
which agrees with the standard ZLK result in the Newtonian limit, $\theta_c=\pm\sqrt{3/5}$.

For small $\epsGR<1$, we find the fit for the two regimes:
\begin{equation}
\left.p(\theta)\right|_{\varepsilon_{{\rm GR}}<1}=\begin{cases}
\frac{p_{c}-0.05(\varepsilon_{\rm GR}+1)}{\theta_{c}^{2}}\theta^2+0.05(\varepsilon_{\rm GR}+1) & |\theta|<\theta_{c}\\
p_{c} & {\rm else}
\end{cases},
\end{equation}
where $p_c(\epsGR) = [0.75 -0.05(\varepsilon_{\rm GR}+1)\theta_c]/(1.5-\theta_c)$, and $\theta_c$ is given by equation \eqref{eqn:thetac}. This fits the large scale features of our distribution reasonably well (Fig.~\ref{fig:fitPlots}).  For larger $\epsGR$, the distribution is flat, so $p(\theta)=0.5$.

\subsection{Argument of Periapsis Distribution}
The distribution of $\omega$ is symmetric about $\omega=\pi$ (Fig.~\ref{fig:fitPlots}), consistent with the first-order dependence of eccentricity variations on $\sin(2\omega)$ according to \cite{Hamers2019SecondOrder}. The shape of the $\omega$ distribution reflects the total precession rate, $\dot{\omega}=\dot{\omega}_{\rm GR}+\dot{\omega}_{\mathrm{tide}}+\dot{\omega}_{\mathrm{ZLK}}$. The ZLK phase portrait has a global minimum of $|\dot{\omega}|$  around $\pi/2$ (both for circulating and librating orbits). In general, finding $|\dot{\omega}|$ requires integration, however, we find we can fit the data well with normal distributions centred at $\omega=\pi/2$ and $\omega=3\pi/2$.

When $\epsGR>30$, rapid precession dominates, yielding a uniform $\omega$ distribution. Otherwise, we find a normal distribution with standard deviation $\sigma_\omega(\epsGR)$ 
\begin{align}
    \sigma_{\omega}=\begin{cases}
0.018\varepsilon_{{\rm GR}}+0.76 & 1\le\varepsilon_{{\rm GR}}\le30\\
-0.58\varepsilon_{{\rm GR}}+1.36 & 0.1\le\varepsilon_{{\rm GR}}\le1
\end{cases}.
\end{align}
When $\epsGR<0.1$, systems undergoing strong ZLK-like oscillations merge before reaching \bMax. Thus the surviving systems with $\epsGR<0.1$ are those with low inclinations, which are not undergoing ZLK-like oscillations, with nearly uniform $\omega$ distributions.  

In the purely Newtonian case, sharp spikes appear at $\omega=\pi/2$ and $3\pi/2$ . These are numerical artefacts caused by systems becoming transiently trapped where $|\dot{\omega}_{\mathrm{ZLK}}|=0$. Such trapping occurs mainly for nearly coplanar systems and is short-lived as angular momentum kicks quickly move them away. This issue vanishes entirely when GR and tidal precession are included. 

The longitude of the ascending node, $\Omega$, is dynamically unimportant for the leading quadrupole order, and remains uniform across all models.

\subsection{Distribution of Stabilized Binaries}

\begin{figure}[]
    \centering
    \includegraphics[width=0.38\textwidth]{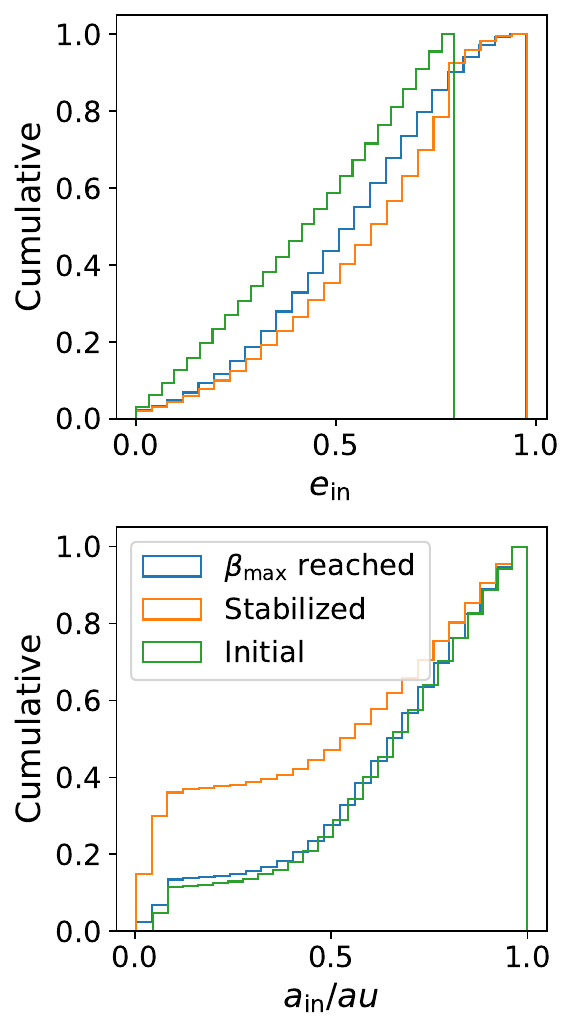}
    \caption{\label{fig:escapedeaDist} Cumulative distributions for systems with different outcomes with both GR and tides turned on. Top panel: eccentricity CDF. Bottom panel: semi-major axis CDF. All systems have $\epsGR<1$. The green lines are the initial distributions. The blue lines are systems that reached \bMax, and the orange lines are those that stabilized.}
\end{figure}
Here, we discuss the distribution of binaries that stabilized (moved far away from the SMBH), focusing on the simulation set that includes both tides and GR. Much like systems that reach \bMax, they initially undergo ZLK-like oscillations, however tidal dissipation circularises highly eccentric binaries, leading to contraction rather than mergers and reducing $\beta$.  This is visible in Fig.~\ref{fig:PercentagePLot}, where with GR only $3\%$ of systems merge and $37\%$ stabilize, whereas including tides changes this to  $0.4\%$ and $45\%$ respectively. Eventually, diffusion will kick the binary away from the SMBH, decreasing $E_{\rm out}$ and increasing $\epsGR$. 

We find that $17\%$ of stabilized systems have contracted for $\epsGR<1$, in agreement with \citet{dodici25}. This is evident from the bottom panel of Fig.~\ref{fig:escapedeaDist}, where the fraction of tighter binaries is larger than the initial conditions. The contracted systems have a peak near $e_{\rm in}=0.8$ (top panel of Fig.~\ref{fig:escapedeaDist}). This is most likely an artefact of our tidal prescription where dynamical tides rapidly shrink the eccentricity to $e_{\rm in}=0.8$ but the equilibrium tides that take over at lower eccentricities are far less efficient. 

The argument of periapsis $\omega$ follows a uniform distribution as the initial sample. GR apsidal precession dominates over ZLK-like precession, erasing any structure and leaving a uniform distribution $\omega$ among binaries that stabilize.

The distribution of $\theta$ is also uniform. For systems that reached \bMax, the non-uniformity in $\theta$ was caused by highly inclined systems that contracted, and thus never reached high $\beta$ values. In contrast, systems that stabilize do so regardless of whether they contracted. Consequently, their $\theta$ distribution remains close to the initial uniform distribution.  

\section{Discussion}\label{sec:Discussion}
\subsection{Comparison to Other Studies}

Our approach is complementary to \citet{Bradnick2017-ed}, but with a methodology more similar to \citet{huangLu2025}. \citet{Bradnick2017-ed} used three-body scattering experiments to explore binaries whose centre of mass is perturbed onto highly eccentric orbits by two-body relaxation (starting with $\beta=0.2$ and evolving toward $\beta \gtrsim 1$).  They ignored GR precession and used only equilibrium tides without accounting for stellar rotation, whereas we employ pseudo-synchronous stellar spins. Meanwhile, we allow energy kicks at periapsis in addition to angular momentum kicks at apoapsis, which are isotropic in 2D rather than 3D. We also investigate the perturbative regime with $\beta < 0.15$ where the second order perturbative approach of \citet{Hamers2019SecondOrder} is valid.  The inner orbit's eccentricity could change by order unity over one outer orbit near $\beta \sim 1$ in the regime examined by \citet{Bradnick2017-ed}, with limited impact from secular effects such as GR precession and ZLK oscillations. We find that the properties of systems that reach \bMax\ are shifted from the initial conditions used by \citet{Bradnick2017-ed}, which will impact their future evolution, including the predicted velocities of hypervelocity stars 
\citep{Rossi2014-re}. 

While the work on this project was being completed, several studies appeared on binary evolution in dense environments:

\cite{Ginat2025-ri} studied the evolution of quasi-hierarchical triples $(\alpha \equiv a_{\rm in}/A_{\rm out} \ll 1)$ with a sufficiently eccentric outer orbit ($1-E_0 \le \sqrt{\alpha}$).\footnote{The orbit still needs to satisfy dynamical stability, $1-E_0 \ge \alpha$.} In this case, the system can behave stochastically. Furthermore, \cite{Ginat2025-ri} have implemented our approach discussed in sec.~\ref{sec:Methods} to preserve the orthogonality relations (Y. B. Ginat, private communication)\footnote{The original preprint of \cite{Ginat2025-ri} showed drift in the eccentricity, analogous to our Fig. \ref{fig:hammersCorrection}. We have informed the authors about the issue and suggested a fix, which was implemented in the accepted version of the paper.}. Although \cite{Ginat2025-ri} focused on the case of $a_{\rm in}/A_{\rm out}=10^{-2}$, in our case $\alpha \lesssim {\rm au\ /\ pc} = 5\times 10^{-6}$ for the widest binaries. This dramatically decreases the fraction of simulated systems that fall into the regime explored by \citet{Ginat2025-ri} if the eccentricity of outer orbits follows a thermal distribution. The fraction of systems in the quasi hierarchical regime is 
\begin{align}
    F(\sqrt{\alpha}<1-E_0<\alpha) & = (1-\alpha)^{2}-(1-\sqrt{\alpha})^{2} \nonumber \\
    & = 2\sqrt{\alpha}-3\alpha+\alpha^{2},
\end{align}
which indeed gives $17\%$ for $\alpha=0.01$ as in \cite{Ginat2025-ri}, but is below $0.4\%$ for our case. Thus the fraction of systems in the quasi-hierarchical regime is negligible.

\cite{huangLu2025} also study the effect of binaries in the galactic center using a similar approach for binary evolution, including GR precession, stellar tides and stochastic two-body relaxation. They explore a wider range of $\beta$ values (from $\beta=0.014$ up to $\beta\geq 1$), more compatible with \cite{Bradnick2017-ed}'s work. \cite{huangLu2025} perform Monte Carlo simulations using discrete sets of initial parameters, and find up to $50\%$ of wide binaries ($a_\inv\sim 1 \ \rm au$) evolve into close binaries through tidal effects due to chaotic tides. Contrary to \cite{huangLu2025}, we draw the initial conditions from \cite{MoeStefano2017}, rather than using discrete parameter sets. \cite{huangLu2025} considered only equal-mass binaries with two zero-age-main-sequence (ZAMS) stars of mass $m_*=0.5\MSUN$ and radius $R_*=0.478\RSUN$. In contrast, we allow for unequal masses and consider systems with stellar masses between $0.1\MSUN$ and $15\MSUN$. We also separate the main physical processes to identify the regimes where each effect is important. Finally, to our knowledge \cite{huangLu2025} chain together SMBH passages using equations from \cite{Hamers2019SecondOrder} without any corrections, which leads to an artificial drift in the inner binary's eccentricity as we have shown, and which may therefore artificially increase the fraction on contracting binaries they find.

\cite{dodici25} study a set-up that includes GR precession, chaotic tides and vector resonant-relaxation (VRR) \citep{Rauch1996-us}, but instead focus on circular outer orbits. They find that roughly one in five binaries originating inside $1\ \rm pc$ contract. \cite{dodici25} also focused on massive B stars and shorter timescales of a few Myr, where VRR, tides, evaporation and ZLK can be important, but neglect the long-term effect of two-body relaxation.

\citet{marklund2025} include two-body relaxation and treat close encounters with direct N-body integrations. They show that stellar flybys play an important role and find higher merger fractions ($29\%$ to $46\%$, depending on $A_\outv$). This is in contrast with \cite{dodici25}, who use a more aggressive version of chaotic tides that effectively suppresses almost all collisions, while \cite{marklund2025} used the model of \citet{mardling1, Mardling1995-vu}, which has a higher transition eccentricity of about $0.8$. This demonstrates the need to include flyby interactions and more detailed tidal models in future work. \citet{marklund2025} also assume that ZLK oscillations are fully suppressed once $\epsGR>1$, whereas we find that significant oscillations can persist up to $\epsGR\simeq 7$. Finally, they switch between secular processes based on their associated timescales, while our results suggest that these processes can act simultaneously rather than sequentially.

\subsection{Limitations and Future Work}
Our model relies on several simplifying assumptions. 

(1) We sample a minority of outer orbits: Only a small fraction of outer orbits will have $E_{\rm out}>0.97$. For a thermal eccentricity distribution, this fraction is $\approx 2 R_{\rm TS}/(\beta_*A_{\rm out})$. For $a_{\rm in} = 1\ \rm au$, $A_{\rm out} = 1\ \rm pc$ and $\beta_*=0.05$, this fraction $2\%$.
Our work is thus valid for the intermediate regime, where the outer orbit is perturbed to a large eccentricity, but with $\beta\ll 1$. Other studies have focussed on the $\beta \sim 1$ regime where the binary is disrupted \citep{Bradnick2017-ed, huangLu2025}). This is different from \cite{marklund2025} and \cite{dodici25}, who focus on circular or at most moderately eccentric outer orbits far from the disruption limit, and study the fates of 'typical' inner binaries. 

(2) Binary evaporation, hardening and strong encounters are not modelled. Soft and semi-soft (near the soft-hard limit) binaries can merge or contract due to flybys \citep{marklund2025}. Tight binaries $a\lesssim 0.06\ \rm au$ can also be dominated by tidal friction and remain stable in otherwise inaccessible environments where all binaries are formally soft \citep[][in prep.]{gri26}. Including these interactions explicitly would likely increase both contraction and merger rates \citep{marklund2025}. 

(3) We assume that systems with $e_\inv<0.8$ evolve under equilibrium tides. Massive stars above $1.2\MSUN$ do not have convective envelopes and evolve via much less efficient dynamical tides raised from internal gravity waves \citep[e.g.,][]{kapil26}. Detailed tidal models will impact the merger and contraction rates of more massive stars.  However, they will not change the parameter distribution of stars that reach \bMax. We also do not account for stellar evolution which could be relevant for the more massive stars that evolve after $\sim 10^8-10^9$ years. These could contribute to mass transfer and post main-sequence stellar mergers.

(4) We define "merger" as the point at which a star overflows its Roche lobe. In reality, Roche lobe overflow can produce stable or unstable mass transfer, including common-envelope evolution, and does not always lead to coalescence. 

(5) We do not account for RR processes. Scalar RR changes the outer angular momentum on timescale $T_{\rm sRR}=M_\bullet^2P_{\rm out}^2/Nm^2T_{\rm coh}$, where $m$ is the typical mass of the stars in the nuclear star cluster, $N$ is the number of stars internal to $A_{\rm out}$, $P_{\rm out}$ is the outer orbital period and $T_{\rm coh}$ is the coherence time. 
In the inner zone $T_{\rm coh}$ is dominated by GR precession $T_{\rm coh}=P_{\rm out} A_{\rm out}/R_g$ where $R_g=GM/c^2$ is the gravitational radius \citep{bar-or16}. For the outer zone $T_{\rm coh}$ is dominated by the mass-precession timescale $T_{\rm coh}= P_{\rm out} M_\bullet/(Nm)$ and the scalar RR timescale is $T_{\rm sRR}=P_{\rm out}M/m$. The scalar RR timescale exceeds that of two-body relaxation at distances greater than $A_{\rm out} \gtrsim 0.06\ \rm pc$ (see Fig. 3 in \citealp{alexander17} and Fig. 4 in \citealp{baror18}). Most outer orbits are drawn near the sphere of influence, neglecting the scalar RR is justified. Vector RR only changes the inclination of the outer orbit, but operates on shorter timescales (but still ZLK oscillations tend to be faster, see \citealp{dodici25}). This may increase the number of binaries undergoing ZLK oscillations and may lead to enhanced contraction or collision rates.

Future work should address these limitations. In particular, to properly model tidal separations, hypervelocity stars, and double TDEs, binaries must be numerically evolved from $\beta=0.15$ up to $\beta=1$. This regime likely requires direct three-body scattering experiments as in \cite{Bradnick2017-ed, huangLu2025}, but using the results of this paper as initial conditions. Finally, in this study we only consider binary stars in the empty loss cone; however, it is also important to consider contributions from binaries in the full loss cone \citep{Verberne2025-bk}.

\section{Conclusion} 

\label{sec:Conclusion}
We have developed and applied a semi-analytical model to investigate the long-term evolution of $10^5$ stellar binaries in the low-$\beta$ regime, incorporating GR precession, stellar tides and angular momentum diffusion. Our approach is based on the analytical approximations for changes in eccentricity and angular momentum \citep{Hamers2019SecondOrder}, but with an additional prescription to preserve the orthogonality relation between the eccentricity and angular momentum vectors, which is not satisfied a priori. This allows us to correct an artificial drift which leads to higher inner eccentricities. 

This enables statistical studies of binary populations close to SMBHs for $10^5$ outer orbits ($\sim 1\ \rm Gyr$). In contrast to previous high-$\beta$ studies that focused on disruption and ejection events, we characterise the perturbative, pre-disruption regime, where long-term secular evolution dominates. We find that GR precession plays the most significant role in shaping the final binary population. It suppresses ZLK-like oscillations, drastically reduces merger rates and determines which systems reach higher $\beta$. Tidal interactions further circularise a subset of binaries but do not strongly alter the overall distributions. 

We provide fitted probability distributions for the final eccentricity, semi-major axis, and orientations of systems that reach $\bMax=0.15$. These distributions can be used as inputs for populations synthesis models or for estimating rates of mergers and TDEs in galactic nuclei. Future extensions include coupling our semi-analytical model with direct N-body integrations to follow binaries from the weak-encounter regime through to deep interactions ($\beta>1$), similarly to \cite{huangLu2025}. Additionally, incorporating the effects of stellar encounters on the inner binary \citep{marklund2025} may provide a more accurate description for wider binaries. 

\section*{Acknowledgements}
We thank Yonadav Barry Ginat for stimulating discussions. We acknowledge support from the Australian Research Council (ARC) center of Excellence for Gravitational Wave Discovery (OzGrav), through project number CE230100016. EG acknowledges support from the ARC Discovery Early Career Research Award (DECRA) DE260101802. YLs research on this subject has been supported by the Simons Investigator grant PG012519.

\bibliography{bibliography}{}
\bibliographystyle{aasjournalv7}

\appendix 
\section{Analytical changes in the eccentricity and angular momentum} \label{appendix:hamers}

In the parabolic limit ($E_{{\rm out}}=1),$ the first and second
order coefficients $\boldsymbol{f}_{e}, \boldsymbol{f}_{j},\boldsymbol{g}_{e}$ and $\boldsymbol{g}_{j}$
are given by Eqns 9 and 27 of \cite{Hamers2019SecondOrder}: 

\begin{align}
\boldsymbol{f}_{e} & =\frac{3\pi}{2}\begin{pmatrix}-3e_{z}j_{y}-e_{y}j_{z}\\
3e_{z}j_{x}+e_{x}j_{y}\\
2(e_{y}j_{x}-e_{x}j_{y})
\end{pmatrix};\quad\boldsymbol{f}_{j}=\frac{3\pi}{2}\begin{pmatrix}-5e_{y}e_{z}+j_{y}j_{z}\\
5e_{x}e_{z}-j_{x}j_{z}\\
0
\end{pmatrix}
\end{align}
 and
\begin{align}
\boldsymbol{g}_{e} & =\frac{3\pi}{16}\begin{pmatrix}75e_{x}^{2}e_{y}-6\pi\left(e_{x}\left(15e_{z}^{2}-6j_{y}^{2}+j_{z}^{2}\right)+6e_{y}j_{x}j_{x}\right)+50e_{y}^{3}+5e_{y}\left(10e_{z}^{2}+j_{x}^{2}-10(j_{y}^{2}+2j_{z}^{2})\right)+50e_{z}j_{y}j_{z}\nonumber  \\
-75e_{x}^{3}-e_{x}(50e_{y}^{2}+5j_{x}^{2}+36\pi j_{x}j_{y}-150j_{z}^{2})-6\pi e_{y}(15e_{z}^{2}-6j_{x}^{2}+j_{z}^{2})+10j_{x}(5e_{y}j_{y}+e_{z}j_{z})\\
-12\pi e_{z}(5e_{x}^{2}+6e_{y}^{2}-3(j_{z}^{2}+j_{y}^{2}))-10(e_{x}e_{y}e_{z}+15e_{x}j_{y}j_{z}-9e_{y}j_{x}j_{y}+5e_{x}j_{x}j_{y})+24\pi j_{x}(e_{x}j_{x}+e_{y}j_{y})
\end{pmatrix}\\
\boldsymbol{g}_{j} & =\frac{3\pi}{16}\begin{pmatrix}75e_{x}^{2}j_{y}+60\pi e_{x}e_{y}j_{y}-6\pi j_{x}(10e_{y}^{2}+15e_{z}^{2}+j_{z}^{2})-50e_{y}e_{z}j_{z}+50e_{z}^{2}j_{y}+5j_{x}^{2}j_{y}\\
-15e_{x}^{2}(5j_{x}+4\pi j_{y})+10e_{x}(6\pi e_{y}j_{x}+5e_{y}j_{y}+15e_{z}j_{z})-6\pi j_{y}(15e_{z}^{2}+j_{z}^{2})-5j_{x}(10e_{y}^{2}+j_{x}^{2}-2j_{z}^{2})\\
-50e_{x}e_{y}j_{y}-50e_{x}e_{y}j_{z}-50e_{y}e_{z}j_{x}-10j_{x}j_{y}j_{z}
\end{pmatrix}.
\end{align}
Note that for the scalar changes of the eccentricity and angular momentum vectors we have $\Delta e = \hat{\boldsymbol{e}} \cdot \Delta \boldsymbol{e} = (\hat{\boldsymbol{e}} \cdot \boldsymbol{f}_e)\varepsilon_{\rm SA} + (\hat{\boldsymbol{e}} \cdot \boldsymbol{g}_e)\varepsilon_{\rm SA}^2$, which is given in terms of the vector or the orbital elements in Eqn 28 of \cite{Hamers2019SecondOrder}.

\section{Orbit Averaged Equations} \label{appendix:A}

\subsection{Procedure}
Here, we describe how to average a quantity over the outer orbit. Suppose each encounter with a field star changes some property, $X$, of the 3-body system by some amount $x(R, b)$, where $R$ is the distance of binary to the SMBH, and $b$ is the impact parameter of the encounter. In the absence of gravitational focusing, the rate of these encounters is given in equation \eqref{eqn:EncounterRate}. Then, assuming the total change does not significantly change the outer orbit over one orbit, we can average over $x$, obtaining
\begin{align}
    \label{eqn:Averaged1}
    \langle x\rangle &=\int_0^{P_\outv}\int_{b_{\mathrm{min}}}^{b_{\mathrm{max}}} x(R(t),b)\frac{\partial \Gamma}{\partial b} {\rm d}b{\rm d}t.
\end{align}
We can perform a coordinate transformation to the true anomaly $F$ of the outer orbit by noting specific angular momentum is given by $J_\outv=R^2\frac{dF}{dt}$, and also $J_\outv=\sqrt{G\BHM A_\outv(1-E_\outv^2)}$. Thus, 
\begin{equation}
    \frac{{\rm d}t}{{\rm d}F}=\frac{R^2}{\sqrt{G\BHM A_\outv(1-E_\outv^2)}}.
\end{equation}
Using the Bahcall-Wolf density \citep{Bahcall1976-bk}, $n(R)=n_0R_0^{7/4}R^{-7/4}$,  equation \eqref{eqn:EncounterRate}, and the velocity dispersion $\sigma(R)=\sqrt{G\BHM/R}$, we find equation \eqref{eqn:Averaged1} becomes
\begin{align}
\label{eqn:Averaged2}
    \langle x\rangle = \frac{2\pi n_0R_0^{7/4}}{\sqrt{A_\outv(1-E_\outv^2)}} \int_0^{2\pi}\int_{b_{\mathrm{min}}}^{b_{\mathrm{max}}} x(R(F),b)R^{-1/4}b{\rm d}b{\rm d}F.
\end{align}

\subsection{Change in Outer Angular Momentum}
\label{sec:two-bodyAppendix}
Each flyby of a field star changes the outer orbit's specific angular momentum by $\Delta J_\outv(R, b) \approx 2Gm_*R/(b\sigma(R))$, where we assume $\sigma(R)\gg \sqrt{Gm_*/a_\inv}$, so we can ignore gravitational focusing. The system is spherically symmetric, and so the sign of $\Delta J_\outv(R, b)$ is equally likely to be positive or negative. Thus, the first moment is $0$, and so we look at the second moment only. Averaging using \eqref{eqn:Averaged2}, we find:
\begin{equation}
    \langle \Delta J^2\rangle = \frac{8\pi n_0R_0^{7/4}G^2m_*^2\ln(\Lambda)}{G\BHM\sqrt{A_\outv(1-E_\outv^2)}} \int_0^{2\pi}R^{11/4}\rm{d}F,
\end{equation}
where $\ln(\Lambda) = \ln (b_{\mathrm{max}}/b_{\mathrm{min}})$ is the Coulomb logarithm. Using the equation
\begin{equation}
    R(F)=\frac{A_\outv(1-E_\outv^2)}{1+E_\outv\cos(F)},
\end{equation}
we obtain
\begin{equation}
    \langle \Delta J^2\rangle = \frac{8\pi n_0R_0^{7/4}G^2m_*^2\ln(\Lambda) R_a^{9/4}}{G\BHM} \int_0^{2\pi}\frac{(1-E_\outv)^{9/4}}{(1+E_\outv\cos(F))^{11/4}}\rm{d}F,
\end{equation}
where $R_a=A_\outv(1+E_\outv)$ is the outer orbit apoapsis. For high eccentricities the integrand evaluates to $\approx1.8$ and we obtain 
\begin{align}
    \langle \Delta J_\outv^2 \rangle \approx1.3\frac{J_c^2(R_{a,\outv})}{\tau_{\mathrm{relax}}(R_{a,\outv})}P_\outv.
\end{align}

\subsection{Change in Inner Angular Momentum}
\label{sec:innerChangeAppendix}
Here we provide a rough derivation of an upper bound for the change of inner-binary angular momentum due to many weak interactions with field stars. We assume the encounter is distant, so that we can use expressions from \cite{Hamers2019SecondOrder}. They find that for a circular binary, the change in inner specific angular momentum is given by 
\begin{align}
    \Delta j_\inv \sim \epsSA j_\inv.
\end{align}
For equal-mass binaries, $\epsSA=(a/b)^{3/2}(1+E')^{-3/2}$, where $E'$ is the eccentricity of the orbit of the flyby star around the binary's centre of mass, and $b$ is the periapsis distance. For hyperbolic encounters, $E'>1$ and so $\epsSA<(a/b)^{3/2}$. Assuming there are $N_{\rm kick}$ flybys per outer orbit, and that each kick is isotropic, we find the total angular momentum change over one outer orbit is 
\begin{align}
    \left(\frac{\Delta j_\inv}{j_\inv}\right)_{\rm total}< \left(\frac{a_\inv}{b}\right)^{3/2}\sqrt{N_{\rm kick}}.
\end{align}
For our systems of interest, using $N_{\rm kick } = P_\outv\langle\Gamma(b)\rangle$ and $R_p = 100 a_\inv / 0.05$, we find that 
\begin{align}
    \left(\frac{\Delta j_\inv}{j_\inv}\right)_{\rm total}<0.03\left(\frac{a_\inv^{9/8}}{b^{1/2}{\rm au}^{5/8}}\right).
\end{align}

\end{document}